\documentclass[aps,apl,reprint,groupedaddress,showpacs]{revtex4-1}

\usepackage{graphicx}
\usepackage{amsmath}
\usepackage{amsfonts}
\usepackage{amssymb}
\usepackage{amsthm}

\begin{document}
\title{Quantum Turbulence of Bellows-Driven $^4$He Superflow: Decay}

\author{S. Babuin${^1}$,  E. Varga${^{2}}$, W. F. Vinen${^{3}}$, L.~Skrbek${^2}$}

\affiliation{${^1}$Institute of Physics ASCR,  Na Slovance 2, 182
21 Prague, Czech Republic\\ ${^2}$Faculty of Mathematics and
Physics, Charles University, Ke Karlovu 3, 121\,16 Prague, Czech Republic\\
${^3}$School of Physics \& Astronomy, University of Birmingham, Birmingham B15 2TT,  UK}

\date{\today}

\begin{abstract}
We report on studies of quantum turbulence with second-sound in superfluid $^4$He in which the turbulence is generated by the flow of the superfluid component through a wide square channel, the ends of which are plugged with sintered silver superleaks, the flow being generated by compression of a bellows. The superleaks ensure that there is no net flow of the normal fluid. In an earlier paper [Phys. Rev. B, \textbf{86}, 134515 (2012)] we have shown that steady flow of this kind generates a density of vortex lines that is essentially identical with that generated by thermal counterflow, when the average relative velocity between the two fluids is the same. In this paper we report on studies of the temporal decay of the vortex-line density, observed when the bellows is stopped, and we compare the results with those obtained from the temporal decay of thermal counterflow re-measured in the same channel and under the same conditions. In both cases there is an initial fast decay which, for low enough initial line density approaches for a short time the form $t^{-1}$  characteristic of the decay of a random vortex tangle. This is followed at late times by a slower $t^{-3/2}$ decay, characteristic of the decay of large ``quasi-classical'' eddies.  However, in the range of investigated parameters, we observe always in the case of thermal counterflow, and only in a few cases of high steady-state velocity in superflow, an intermediate regime in which the decay either does not proceed monotonically with time or passes through a point of inflexion. This difference, established firmly by our experiments, might represent one essential ingredient for the full theoretical understanding of counterflow turbulence.

\end{abstract}

\maketitle

\section{Introduction}

In this paper we report on an experimental investigation of the decay of turbulence in a quantum fluid, superfluid $^4$He, displaying the two-fluid behavior. The necessary introduction to superfluidity and quantum turbulence is given in the next section, together with a review on counterflow turbulence, a unique form of turbulence existing in superfluid $^4$He, of which the superflow treated in this article represents a special case. The review sets this work into a detailed context. Here we summarize the general motivation for this project, which is two-fold.

(i) We wish to continue the study of temporal decay of turbulence in quantum fluids, in the spirit of exploring similarities and differences with the decay of turbulence in classical viscous fluids, as advocated again as an important goal for the community in the latest review on the state of the field~\cite{PNAS_intro}. The decay of turbulence is indeed a cornerstone in classical turbulence studies because the rate of decay (of kinetic energy or vorticity) is related to the distribution of energy over the scales of the system, which constitutes a key description of a turbulent flow~\cite{TennekesLumley}. In quantum turbulence our experiments can accurately measure the decay of the total length of quantized vortices per unit volume, a well defined quantity which, if the detailed spatial distribution of the vortex lines is known, can be related to classical vorticity.

(ii) Our second and more specific motivation is to continue our investigation of the mechanically-driven turbulent flow of the superfluid component of $^4$He, of which we have reported the steady-flow characteristics in this journal~\cite{babuin_steady_superflow}. The study of pure superflow ought to be in principle simpler than thermal counterflow because the normal component is on average at rest. In our previous work we have demonstrated that Galilean invariance holds between steady-state counterflow and superflow turbulence, i.e. that, to first order, the turbulence produced when the superfluid and normal components of $^4$He move in opposite direction is the same as when the normal fluid is at rest and the superfluid moves past it with the same relative velocity, despite the fact that the presence of a finite channel ought to introduce differences such as a profile to the normal component. In this paper we extend this work by asking how these two turbulent flows decay in time when driving is suddenly switched off. We describe emerging similarities as well as differences, which ought to deepen our general understanding of the underlying physics of quantum turbulence, especially in relation to the dynamical state of the normal component.

\section{Review of counterflow turbulence}
\label{sec:review}

Quantum turbulence~\cite{PNAS_intro, vinen_niemela_rev, skrbek_sreeni_rev, nemirovskii_review} is the turbulence occurring in a superfluid such as the superfluid phases of liquid $^4$He and liquid $^3$He~\cite{tilley}. At a finite temperature superfluids exhibit two-fluid behavior, a normal viscous fluid (composed of thermal excitations) coexisting with an inviscid superfluid component. Flow of the superfluid component is strongly influenced by quantum effects,  reflecting the origin of superfluidity in Bose condensation.  In the case of $^4$He superfluid flow must be irrotational,  rotational motion being possible in a simply-connected volume only with the formation of topological defects in the form of vortex lines,  each of which carries a circulation of $\kappa = h/m \thickapprox 1 \times 10^{-3}$~cm$^{2}$/s,  where $h$ is Planck's constant and $m$ is the mass of a helium atom.  Turbulence in the superfluid component must therefore in general take the form of a complex tangle of vortex lines. A purely random tangle involves turbulent energy almost entirely on only \emph{quantum length scales} -- scales comparable with, or less than, the mean vortex spacing, $\ell = L^{-1/2}$,  where $L$ is vortex line density (length of line per unit volume),  although local polarization of the vortices can lead to the existence of turbulent energy on any larger scale~\cite{vinen_classical, vinen_niemela_rev, skrbek_sreeni_rev}. Both purely quantum and classical features of turbulence can therefore be detected simultaneously in the same quantum flow, depending on the length scale at which this quantum flow is probed~\cite{MarcoEPL,MarcoPRB}.

If  vortex lines move relative to the normal fluid they experience a drag force, called mutual friction~\cite{vinen_niemela_rev, skrbek_sreeni_rev}.
On \emph{quasi-classical length scales}; i.e., scales large compared with $\ell$, the superfluid component usually behaves like a classical fluid at high
Reynolds number.  The small kinematic viscosity of the normal fluid (of order $\kappa/6$~\cite{LvovSkrSreeni}) means that the same is true for the normal fluid on these large quasi-classical scales.  Thus the two fluids can move together with the same velocity fields,  mutual friction serving only to stabilize this coupled motion.  The coupled fluids then behave as a single quasi-classical fluid at high Reynolds number. This situation, referred to as \emph{coflow}, obtains quite frequently; for example when the fluid is stirred by large scales objects similarly as for classical fluids, such as propellers~\cite{maurer}, grids~\cite{skrbek_sreeni_10ch}, flow through channels~\cite{salort, vansciver_coflow, babuin_coflow_decay, Babuin_PF} or due to various oscillating objects~\cite{PLPT}.

On quantum length scales (even in coflow) the superfluid motion is strongly influenced by quantum effects.  The coupled motion therefore cannot be maintained,  the resulting dissipation due to the mutual friction, combined with viscous dissipation in the normal fluid, results in a dissipation per unit mass of helium of the form
\begin{equation}
\epsilon=\nu'\kappa^2 L^2,
\label{eq:diss}
\end{equation}
where $\nu'$ is an effective kinematic viscosity of order $\kappa$~\cite{vinen_niemela_rev, Walmsley_rev, prague_eff_kin, skrbek_sreeni_10ch, LvovSpectra}. This type of coupled motion, leading to quasi-classical behavior on large length scales, will usually obtain as long as there is no forced relative motion between the two fluids on these scales.  We shall refer to such forced relative motion as \textit{counterflow}. It is most easily imposed with a temperature gradient, the superfluid component moving up the gradient and the normal fluid moving down, with no net mass flow. This is one special case of counterflow, known as \textit{thermal counterflow}.

Quite generally, by combining mechanical and thermal drive, a rich variety of counterflows, i.e., two-fluid flows with different flow ratio of the two components, can be generated, representing  a very complex superfluid hydrodynamic system. In this paper, we are concerned with the special case of counterflow that can be generated by forcing helium through a tube, the ends of which are closed with superleaks -- only the superfluid component can flow through the superleaks, so that the average velocity of the normal fluid vanishes. The forced flow is conveniently driven by compression of a bellows,  and we refer to this type of counterflow as \textit{bellows-driven superflow}~\cite{babuin_steady_superflow},  or simply \textit{superflow}.

Early experiments on thermal counterflow~\cite{vinen_steady, *vinen_transient, *vinen_theory, *vinen_crit} led to the idea that in such a flow a self-sustaining random tangle of vortex lines is generated simply by the imposed relative motion,  the turbulence being essentially homogeneous, and a phenomenological equation was derived that describes the growth and decay of line density. In the steady state the line density is given in terms of the relative velocity $v=(v_s-v_n)$  by
\begin{equation}  L=\gamma v^2,
\label{eq:steadyL}
\end{equation}
where $\gamma$ is a temperature-dependent parameter.
Confirmation of these ideas,  together with an understanding of the physical processes involved, came from the pioneering computer simulations of Schwarz~\cite{schwarz1978, *schwarz1983, *schwarz1985, *schwarz1988}, which were improved and refined by Adachi \textit{et al.}~\cite{adachi}, Baggaley and Barenghi~\cite{baggaley_tree} and Kondaurova \textit{et al.}~\cite{Kondaurova} . These simulations were based on the vortex filament model. They assumed that the flow of the normal fluid remained laminar,  and that the mean flow velocities were spatially homogeneous and in an unbounded volume.  The resulting vortex tangle was disordered,  so there was no large scale turbulent motion in the superfluid component.

In Ref.~\cite{babuin_steady_superflow} we reported the results of a study of the attenuation of second-sound in bellows-driven steady-state superflow. According to the theories that we have described so far,  the line densities observed in such an experiment should agree with those measured at the same value of $v$ as those observed in thermal counterflow. This was indeed confirmed to a good approximation in channels of different cross-sections, with the agreement being particularly good when counterflow and superflow are measured in the same (large) channel \cite{Babuin_PF}.

There are, however, subtle differences between counterflow and superflow, even in the steady-state. Experiments have shown that there is a critical velocity below which the line density is unmeasurably small, which scales with the channel size, $D$. It was found to be roughly temperature independent scaling as $D^{-1/4}$ for superflow, while for counterflow it displays a $D^{-1}$ temperature dependent scaling~\cite{babuin_steady_superflow}. Additionally, another critical velocity has been reported  by Tough and his co-workers~\cite{tough_rev} in thermal counterflow, above which the  $\gamma$ factor suddenly increases, the so-called T1-T2 transition. The existence of this transition depends on the details of channel cross-section in counterflow (according to the data reviewed by Tough~\cite{tough_rev} two states are observed if the aspect ratio is of order one \emph{and} the smallest dimension is less than about 1~mm; a single state is observed if the aspect ratio is much larger than 1 \emph{or} the smallest dimension is larger than 1~mm). In superflow on the contrary, in channels of different cross-section, only one regime of turbulence has been observed. These facts may be related to the different dynamics of the normal component in the two systems.

Theoretical and computational work has recognized that earlier simulations were unrealistic in at least one important respect:  they related to an unbounded volume of helium,  and to a situation where the velocity of the normal fluid (assumed laminar) and the mean velocity of the superfluid are spatially uniform.  In practice,  thermal counterflow takes place almost always in a channel of finite cross-section,  so that, at the very least,  the no-slip condition for the viscous normal fluid at a solid boundary must lead to spatial non-uniformity in the velocity of the normal fluid. Furthermore, the Reynolds number for the flowing normal fluid is typically quite large,  so that it is questionable whether the flow of the normal fluid remains laminar.

The results of more realistic theoretical and computational work can be summarized as follows.  The stability of laminar normal-fluid flow has been studied by Melotte and Barenghi~\cite{melotte&barenghi},  although with assumptions that have turned out to be unrealistic (the normal-fluid velocity profile remains approximately parabolic in the presence of mutual friction, and the line density remains unperturbed and spatially uniform). The properties of the vortex tangle have been studied when the normal fluid has a prescribed laminar parabolic profile by Aarts and De Waele~\cite{aarts}, Adachi and coauthors~\cite{adachi} and Baggaley and Laurie~\cite{baggaley_laurie}; and when the normal fluid has imposed on it a prescribed classical turbulent flow profile by Baggaley and Laizet~\cite{BaggaleyLaizet}.  In the former case the line density turns out to be quite inhomogeneous,  although the spatially averaged value of $\gamma$ is not seriously affected. In the latter case the value of $\gamma$ is increased,  suggesting that the transitions at which $\gamma$ is observed to increase are associated with transitions to normal-fluid turbulence. However,  this work is still unrealistic for two reasons:  the motion in the two fluids,  coupled by mutual friction, is not treated in a dynamically self-consistent way;  and it is assumed that there is no pinning of vortex lines at a solid boundary.

We note at this point that our recent experimental work on bellows-driven superflow~\cite{babuin_steady_superflow} showed that the steady-state spatially-averaged line density is hardly any different from that observed in thermal counterflow at the same relative velocity $v$. This is in spite of the fact that the normal fluid velocity,   relative to the channel walls, is very different in the two cases. This suggests that the average line density is insensitive not only to any inhomogeneity in the vortex tangle, as suggested by simulations, but perhaps also, contrary to the simulations, to any turbulence in the normal fluid, since any turbulence in the normal fluid might be expected to be different in the two types of flow.  We are led to conclude that detailed information about counterflow turbulence must come from experiments other than those that measure average line densities in steady-flows: two possible directions are the direct visualization of the turbulence and a study of the decaying line density after the flow is switched off, which we pursue in our work.

On the visualization side, experiments on thermal counterflow have been reported using as tracers both micron-sized hydrogen or deuterium particles and metastable He$_2$ excimer molecules~\cite{PNAS_visualization}. In the former case at relatively low heat currents bimodal distributions of vertical velocity have been first measured by Paoletti and coworkers~\cite{paoletti}, indicating that some of the particles move in the direction of the normal fluid, while others are trapped on vortices and move with the tangle, with a velocity generally different from that of the superfluid. At larger heat currents, where effects relevant to our present studies might be seen, the particles experience frequent trapping and de-trapping: vertical velocity distribution changes to a monovalued one~\cite{chagovets_vansciver} and interpretation is in general harder. A wealth of statistics of particle velocity and acceleration has been produced by the Prague group in counterflow in different heat current regimes~\cite{MarcoJFM, MarcoEPL, MarcoPRB}, showing that the character of particle dynamics changes from classical to quantum as the length scale investigated is reduced from well above to well below the inter-vortex separation. Additionally, there has also been some indication that vortical structures exist on scales larger than the intervortex spacing~\cite{MarcoPRB}. The excimer molecules instead are useful because they track only the normal fluid (at temperatures above 1K). The use of these molecules is still at an early stage of development,  but Guo and coworkers have already produced evidence that the normal fluid does become turbulent above a critical velocity~\cite{guo2010}. Very recently, Guo's group have been studying the time evolution of thin lines of excimer molecules produced in counterflowing helium,  from which they can deduce the flow of the normal fluid in greater detail~\cite{guo2015}. We will further refer to this work in the Discussion session.

We now turn to the main topic of this paper, i.e. the decay of vortex line density in superflow and counterflow.
The phenomenological Vinen equations describing the growth and decay of line density in counterflow turbulence~\cite{vinen_theory} predict that the free decay should be described by the equation
\begin{equation}
\frac{dL}{dt} =  - \frac{\chi_2  \kappa}{2 \pi}  L^2,
\label{eq:Vinen_decay}
\end{equation}
where $\chi_2$ is a dimensionless temperature-dependent parameter proportional to mutual friction~\cite{schwarz1988, LvovPRBR},  related closely to the effective kinematic viscosity $\nu'$.  It follows that
\begin{equation}
L = \Big[\frac{1}{L_0} + \Big(\frac{\chi_2  \kappa}{2 \pi}\Big) t\Big]^{-1},
\label{eq:Vinen_decay_solution}
\end{equation}
where $L_0$ is the line density at time $t=0$.  Recent simulations of Mineda \emph{et al.} based on the assumption that counterflow turbulence is homogeneous and occurs in an unbounded medium with laminar flow of the normal fluid have confirmed this prediction~\cite{mineda}. However, it is now well-known~\cite{vinen_transient, schwarz&rozen, schwarz&rozenPRL, prague_counter1, prague_counter2, barenghi_skrbek_rev}, that the experimentally observed decay is quite different.  Although there is for a very short time a rapid decay that might be described by Eq.~(\ref{eq:Vinen_decay_solution}),  the decay then slows down for a time and may actually reverse (there is a ``bump" in the decay curve),  while at large times the decay is observed to continue as $t^{-3/2}$ rather than as $t^{-1}$. Note that in the rest of this paper we shall use the jargon word ``bump'' to refer to both the case of actual increase in vortex line density during decay (i.e. change of sing in $dL/dt$) and also to the simple slow down in decay rate associated to the presence of a point of inflexion (i.e. a change of sing in $d^2L/dt^2$), and we will be more specific if needed.

It is now widely accepted that the time-dependence as $t^{-3/2}$ is associated with the decay through a quasi-classical Richardson-Kolmogorov cascade of coupled (superfluid-normal fluid) energy containing eddies,  the size of which is determined by and limited by the dimensions of the containing channel~\cite{vinen_classical, skrbek_stalp, skrbek_sreeni_rev, skrbek_sreeni_10ch}. It has been shown that this decay is described in detail by the equation
\begin{equation}
L(t) = \frac{D(3C)^{3/2}}{2\pi \kappa \nu'^{1/2}}(t - t_0)^{-3/2},
\label{eq:classical_limit}
\end{equation}
where $D$ is taken as the channel width, $C \approx 1.5$ is the classical Kolmogorov-41 constant, and $t_0$ denotes the virtual time origin.

Behaviour described by Eq.~(\ref{eq:classical_limit}) was first observed by the Oregon group in the decay of grid turbulence~\cite{smithOregon, stalp, skrbek_oregon}, where the formation of large eddies can be understood in classical terms.  However, formation of large-scale classical eddies in the decay of counterflow turbulence must be less straightforward than is the case with grid turbulence.  Are they formed out of large scale eddies already present in the steady state?  Or are they generated from scratch during the early stages of the decay?  Presumably,  the processes involved in this formation are reflected in the early stages of the decay of line density,  but interpretation is hard. There has been some speculation about these early stages,  particularly about the origin of the ``bump''~\cite{barenghi_skrbek_depol}, to which we shall refer later, but there is as yet no agreed explanation.  Further substantial progress  must probably await the results of visualization experiments, perhaps backed up by more realistic numerical simulations. However, we argue that if indeed the large eddies responsible for the $t^{-3/2}$ decay are formed out large-scale eddies present in the steady state,  then bellows-driven superflow could exhibit very different features in its decay from that seen with thermal counterflow. We have therefore undertaken a study of the decay of such bellow-driven superflow with a direct comparison of thermal counterflow studied in the the same channel and conditions, and we present the results in this paper.  As we shall see, the two cases do indeed exhibit different forms of decay.

\section{The experiment}
\label{sec:experiment}
\subsection{Apparatus and method}
\label{subsec:apparatus&method}

\begin{figure}
\includegraphics[width = 1\linewidth]{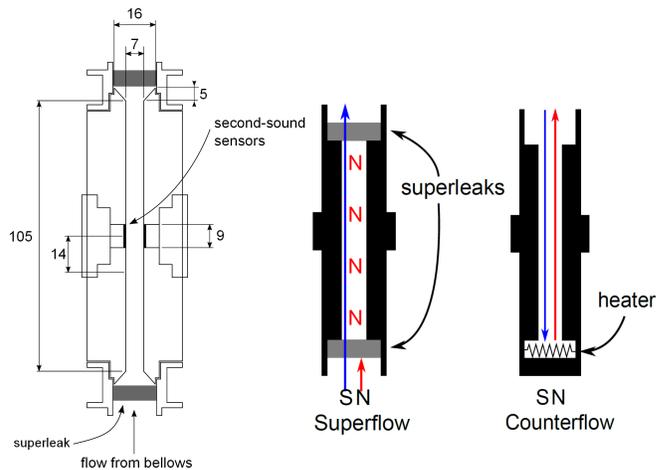}%
\caption{\label{fig:setup}(Color online) Flow channel for turbulent superflow decay studies, to scale, with dimensions in mm (left). Helium flows vertically up, provided by low temperature bellows next to the channel. Superleak filters allow flow-through of superfluid component only. Second-sound probes measure the total length of quantized vortices per unit volume, Eq.~\eqref{eq:L(t)}. The channel can be modified to host thermal counterflow too, as illustrated by sketches on the right, where S and N stand for superfluid and normal components of He-II. The channel exists in two variants, with internal square cross-section 7~mm and 10~mm in side, coded D7 and D10 in the article.}
\end{figure}

A full description of the mechanically driven superflow apparatus and the measurement technique was given in Ref.~\cite{babuin_steady_superflow}. Here we briefly recall only the essential features for convenience, focussing on the aspects more specific to the decay studies. A drawing of the flow channel is in Fig.~\ref{fig:setup}. Two vertical brass flow channels have been used in this experiment. The test section is 105~mm in length and has an internal square cross-section of side 7~mm and 10~mm, therefore with a factor 2 change in cross-sectional area (coded D7 and D10 in this article). The channel ends are plugged by sintered-silver superleak disks, each 2~mm thick, 16~mm diameter, with 1/2 filling fraction; these discs serve to prevent any net flow of the viscous normal component on time scales relevant to the experiment.

The superflow is driven by a low temperature stainless steel bellows immersed in the open cryostat bath and operated through a shaft by a computer-controlled precision motor located above the cryostat at room temperature. The velocity of the flow in the channel is inferred from a measurement of the rate of compression of the calibrated bellows volume, assuming that the helium is incompressible. Counterflow is studied in the traditional way, as in previous Prague experiments~\cite{prague_counter1, prague_counter2}: we used the same channel of the superflow experiment (D10 only), installed it horizontally in the cryostat, removed one superleak to open the channel to the bath, and fitted the other end with a close cap which hosts a 50~$\Omega$ heater resistor (sketch in Fig.~\ref{fig:setup}). The dissipated power is measured continuously, by independent measure of voltage and current.

Turbulence is detected by measuring the extra attenuation of second-sound caused by scattering of normal-component thermal excitations by the vortex lines. Second-sound is generated and detected by a pair of vibrating porous membranes located in the walls of the channel at its mid-point (see Fig.~\ref{fig:setup}); the second-sound travels across the channel, which acts as a resonator.
The time-dependent attenuated amplitude of second-sound at resonance $a(t)$ can be related to the instantaneous total length of quantized vortex lines per unit volume, $L(t)$ through the equation:
\begin{equation}
L(t) = \frac{6\pi \Delta f_{0}}{B \kappa}\left(\frac{a_{0}}{a(t)} - 1 \right),
\label{eq:L(t)}
\end{equation}
where $a_0$ and $\Delta f_{0}$ are the amplitude and full width at half maximum of the second-sound amplitude resonant curve for quiescent helium, and $B$ is the mutual friction coefficient of order unity, tabulated in Ref.~\cite{donnelly_barenghi_tables} (the frequency dependence of $B$ can be neglected in this experiment since we perform measurements only with a single low frequency mode). The attenuation of second-sound measures the length of vortex line per unit volume weighted by a factor $\sin^2 \theta$, where $\theta$ is the angle between  any element of vortex line and the direction of propagation of the second-sound. To derive Eq.~\eqref{eq:L(t)}~\cite{babuin_steady_superflow}, the distribution of vortex lines is assumed to be fully random and spatially uniform; the opposite limiting case where the lines are instead assumed to be fully polarized, i.e. forming vortex rings lying in planes perpendicular to the flow direction, leads to a version of Eq.~\eqref{eq:L(t)} a factor 4/3 higher. Therefore if the real vortex line distribution is not known, the use of Eq.~\eqref{eq:L(t)} can lead to errors in $L(t)$ of at most 33\%. We will return to how this aspect may affect results during discussion.
\begin{figure}
\includegraphics[width = 1\linewidth]{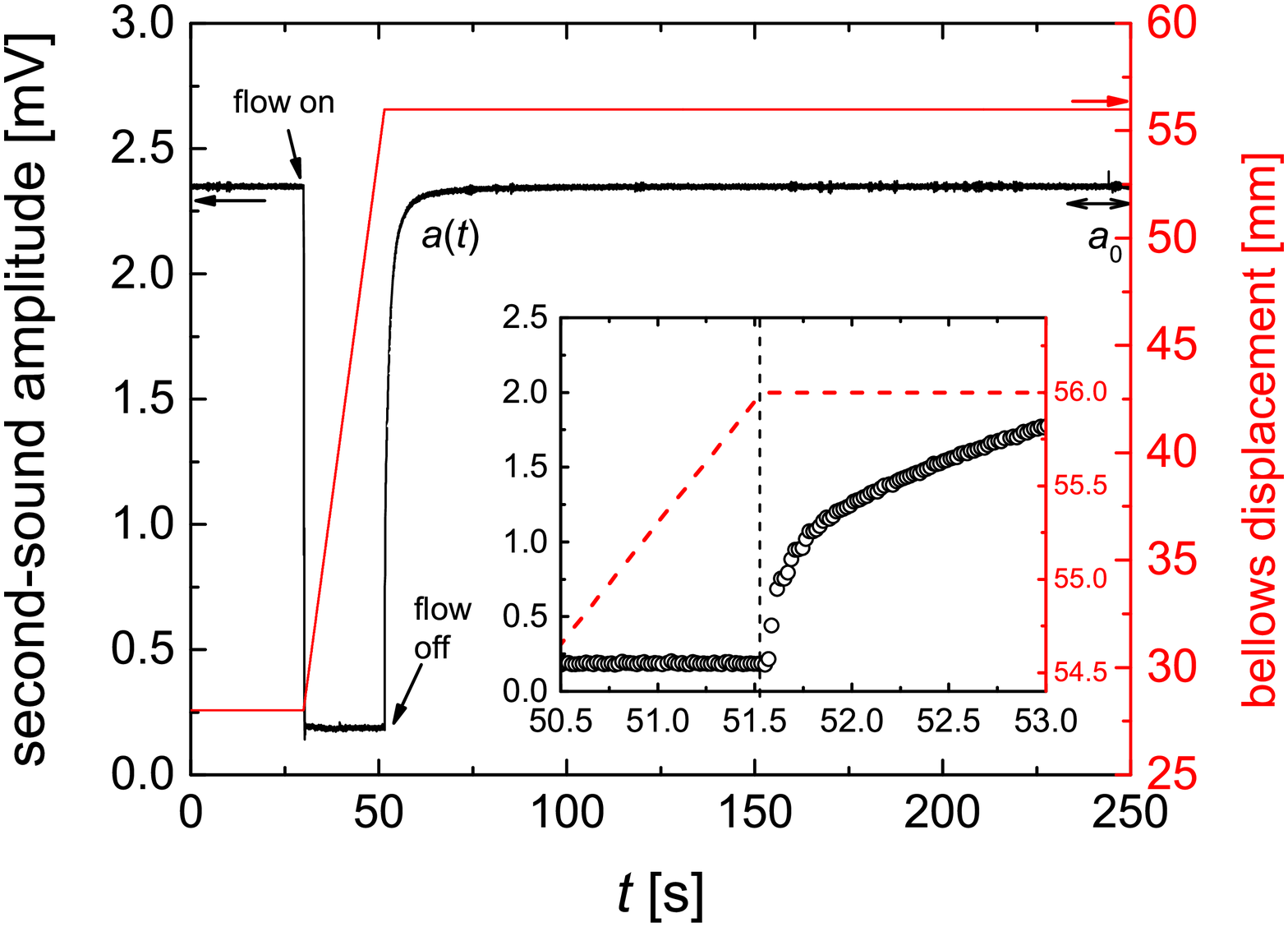}%
\caption{\label{fig:rawdata}(Color online) Example of a single raw measurement of superflow turbulence decay in  D7 channel, at T = 1.26~K. The second-sound amplitude $a(t)$ (black solid line) is plotted as a function of time. At $t =0$ helium is quiescent, at $t = 30$~s a steady-flow of velocity 11.3 cm/s is produced by the bellows, and at $t = 51.5$~s it is suddenly switched off, resulting in the following decay. The reference signal $a_0$ entering the calculation of vortex line density, Eq.~\eqref{eq:L(t)}, is the average of the last 20 s of the amplitude signal. The red solid line (right axis) shows the bellows displacement. Inset: a zoom around the switch-off time, showing the second-sound reacts immediately after the halting of the drive. Time resolution of our measurements is about 16~ms.}
\end{figure}

The decay process is too fast in time to allow sampling of a full resonance curve at any point during the decay.  However, our studies of the steady-state have shown that the second-sound resonant frequency is not significantly affected by the extra attenuation, so that it is sufficient to set the second-sound frequency on resonance and observe this resonant response  as a function of time, as shown in Fig.~\ref{fig:rawdata}. This response accurately reflects the changing attenuation only if the natural response time of the second-sound resonator (of order the inverse line-width) is sufficiently short; in practice $\Delta f^{-1}$ is about 10~ms at the start of the decay and about 100~ms for quiescent helium, values that are small enough for our purposes. Amplitude times-series are sampled at 60~Hz, and each sample is averaged with a lock-in amplifier with time constant of 10~ms. Temperature control in our cryostat is of order 0.1~mK, therefore sufficient to ensure that temperature drift cannot cause significant drift away from resonance, as we have experimentally verified.

To create a turbulent steady-state, the bellows are compressed at a constant rate for about 20~s (Fig.~\ref{fig:rawdata}), after which compression is suddenly stopped. The position of the movable end of the bellows is recorded by the motor encoder,  which shows that sudden stopping is achieved in less than 10 ms (inset). The decay of line density is followed for 200~s, after which the amplitude of the second-sound signal has reached a statistically steady value; a further 30 s is allowed to elapse before a new measurement is made. The second-sound amplitude before and long after the steady flow was observed to be generally slightly different. This effect -- related to a varying remanent amount of vortex line in the sample -- is studied statistically in Section~\ref{subsec:remanent} and has some bearing on the interpretation of results. It is not clear whether the beginning or the end of the amplitude time series ought to be used for $a_0$ in Eq.~\eqref{eq:L(t)}: we calculate it averaging the last 20~s. The measurement protocol for thermal counterflow decays is very similar.

The parameter space covered by the experiment is summarized in Table~\ref{tab:parameter_space}. We have performed mechanically-driven superflow and thermally
driven counterflow decay measurements, in the two channels D7 and D10, in the temperature range between $1.25~\rm{K}$ to $2.10~\rm{K}$, and for initial
steady-state line densities, $L_0$, spaced almost exactly one decade apart: $10^6$~cm$^{-2}$, $10^5$~cm$^{-2}$ and $10^4$~cm$^{-2}$ (we shall refer to the
decay curves corresponding to these initial line densities as $L6$, $L5$ and $L4$). The table shows the corresponding superflow and counterflow velocity in
the steady-state, with experimental uncertainty of 3\%.
\begin{table}
\centering
\begin{tabular}{|c|c|c|c|c|}
\hline
& & \multicolumn{2}{|c|}{Mechanical superflow} & Thermal counterflow \\
\hline
 &  & D7 &  D10 & D10\\
\hline
$T$[K] &  $L_0$ [cm$^{-2}$] & $v_s$ [cm/s] &  $v_s$ [cm/s] &  $v_{ns}$ [cm/s]\\
\hline
 & $10^{6}$ & 11.3 &  / & /\\
1.25 & $10^{5}$ & 3.6 &  4.25 &/ \\
 & $10^{4}$  &  / & 1.3 & /\\
\hline
 & $10^{6}$ & 7.7 &  7.4 & 8.45\\
 1.45 & $10^{5}$ & 2.6  & 2.7 & 3.0\\
  & $10^{4}$ & 0.86  & 1.0 & 1.0\\
\hline
& $10^{6}$ & 5.7  & 5.3 & 7.1\\
 1.65 & $10^{5}$ & 1.9  & 2.0 & 2.5\\
& $10^{4}$ & 0.57 & 1.0 & 0.9\\
\hline
& $10^{6}$ & 4.9 & / & / \\
1.75 & $10^{5}$ & 1.7 & / & / \\
  & $10^{4}$ & 0.6 &  / & / \\
\hline
& / & / & / & / \\
2.10 & $10^{5}$ & 0.4 & / & 1.50 \\
  & $10^{4}$ & 0.15 & / & 0.51 \\
\hline
\end{tabular}
\caption{Overview of the parameter space explored by the experiment. Steady-state line density and corresponding mean superflow and counterflow velocity, shown per temperature and channel width (7~mm and 10~mm side of square cross-section).}
\label{tab:parameter_space}
\end{table}
\begin{figure}
\includegraphics[width = 1\linewidth]{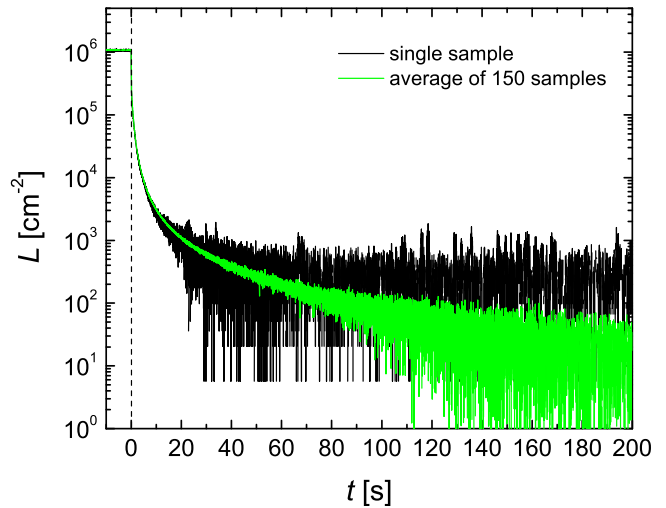}%
\caption{\label{fig:ensemble}(Color online). Demonstration of the improvement of signal-to-noise ratio by ensemble average of 150 decay samples. Data relates to pure superflow, D7, T = 1.26~K, L6. The noisier signal in solid black line is a single sample randomly chosen from the batch of 150. Time is rescaled here, with $t = 0$ marking the instant when the bellows stops. On the y-axis is the vortex line density calculated from the data in Fig.~\ref{fig:rawdata} via Eq.~\eqref{eq:L(t)}. The average signal only looks noisy at late times because it averages to zero and is plotted in logarithmic scale. We can resolve 6 orders of magnitude in decay of vortex line density.
}
\end{figure}

For every combination of temperature and starting line density we have measured typically 150 individual decays, under nominally identical experimental conditions and we have ensemble-averaged these samples, by linearly interpolating each one onto a 100 Hz time-series and averaging point-wise. Decay signals are checked individually and rare anomalous ones are discarded from averaging. The improvement of the averaged signal with respect to a single sample is demonstrated in Fig.~\ref{fig:ensemble}. Averaging over a large ensemble has proven essential for our study, allowing us to resolve 6 orders of magnitude of decay on $L$.

\subsection{The remanent vortex line density}
\label{subsec:remanent}
\begin{figure}
\includegraphics[width = 1\linewidth]{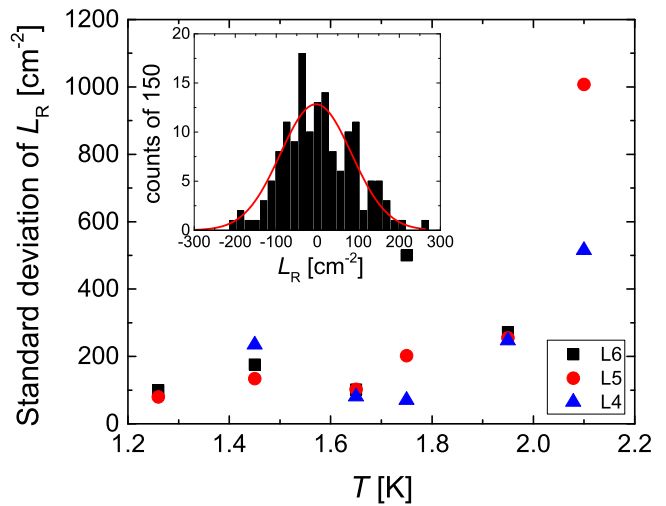}%
\caption{\label{fig:statistics}(Color online) The dependence of the standard deviation of the remanent vortex line density $L_R$ (vortex lines remaining in the channel after a decay process) with temperature and steady-state line density. Data relates to the D7 channel, D10 being similar. $L_R$ is calculated with Eq.~\eqref{eq:LR} and its standard deviation is typically from a batch of 150 decays. The distribution of $L_R$ in a typical batch (here L5 at $T = 1.65$~K) is shows inset, roughly with a Gaussian distribution with zero mean.}
\end{figure}

\begin{figure*}
\includegraphics[width = 0.90\linewidth]{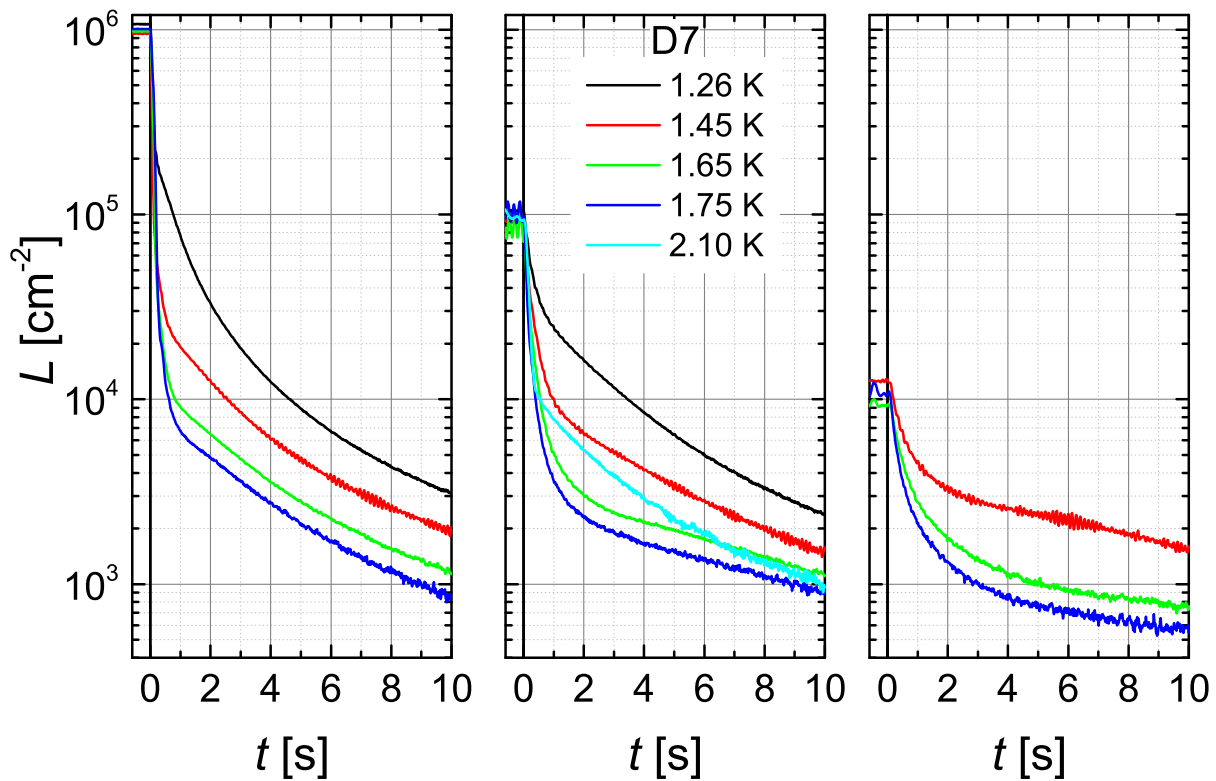}%
\caption{(Color online) Decay of vortex line density of pure superflow turbulence. Plots of $L$ (in log scale) \textit{versus} $t$ for the D7 channel, at different temperatures. Each panel groups decays from the same steady-state density, at different temperatures. A version of this Figure in log-log coordinates is in the Supplemental Material (SM).}
\label{fig:sf_D7_short}
\end{figure*}

\begin{figure*}
\includegraphics[width = 0.90\linewidth]{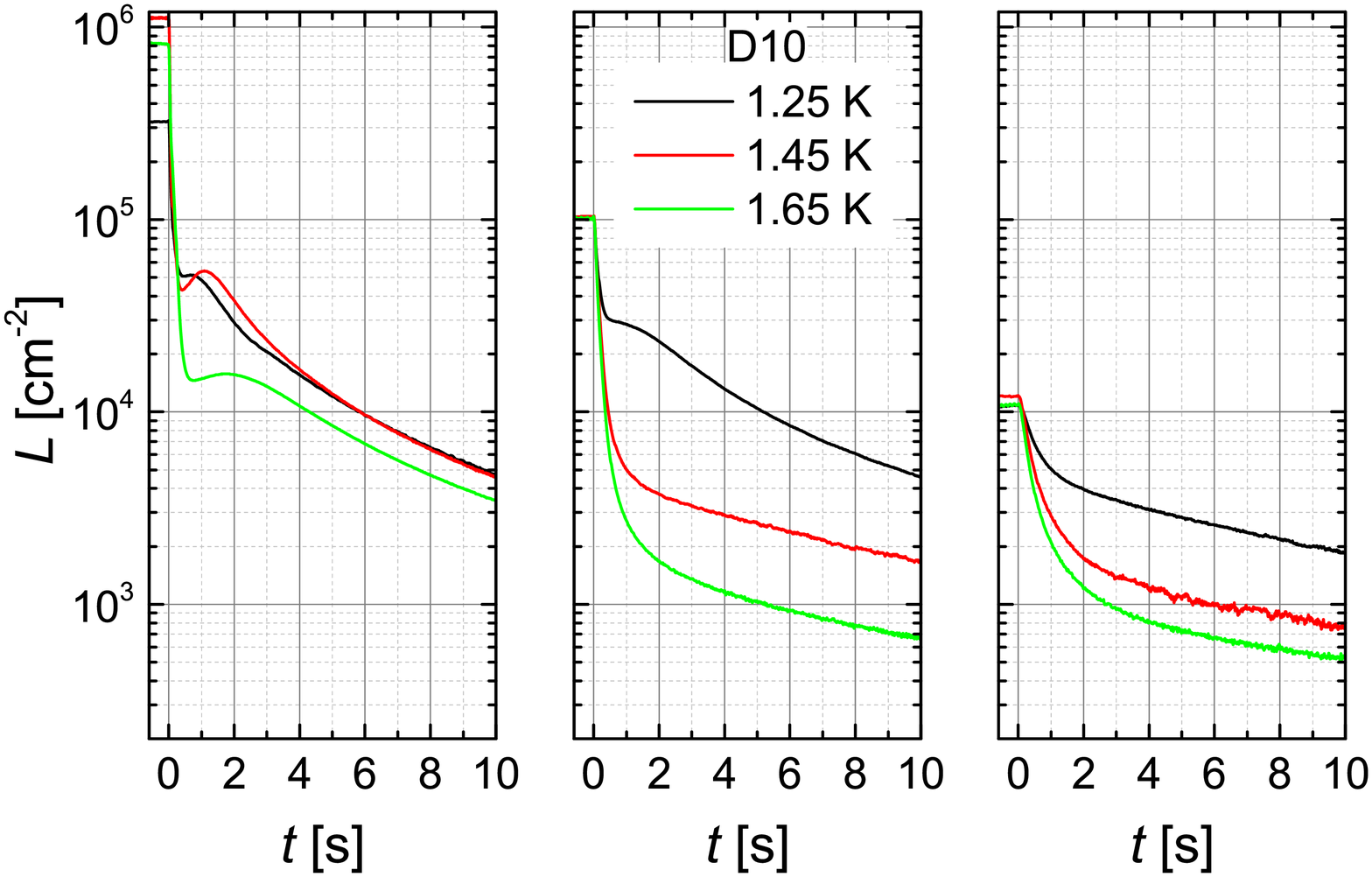}%
\caption{(Color online) Decay of vortex line density of pure superflow turbulence, as in Fig.~\ref{fig:sf_D7_short} but for channel D10. Features are similar to D7, except that L6 curves display a short non-monotonic behavior and the L5 curve at 1.25~K has a change in the decay rate with a point of inflexion. This feature, referred to in the text as a ``bump'', is an exception in superflow decay and the rule in counterflow decay, shown in later figures.}
\label{fig:sf_D10_short}
\end{figure*}

Our second-sound attenuation technique cannot provide an absolute measurement of vortex line density, because the reference second-sound signal $a_0$ in Eq.~\eqref{eq:L(t)} may itself be attenuated by remanent vortex lines persisting in quiescent helium after the turbulence has decayed. Vortex lines are indeed expected to survive because they pin effectively to any surface, due to their {\AA}-size core~\cite{PNAS_intro}. The absolute vortex line density in our sample is therefore $L_{\mathrm{abs}} = L_{\mathrm{rem}} + L$, where $L_{\mathrm{rem}}$ is the non-measurable remanent density of vortex lines hidden in $a_0$. However, $L_{\mathrm{rem}}$ is expected to be negligibly small compared to $L$ during most of the decay. An estimate provided by the work of Awschalom and Schwarz~\cite{awschaloms&schwarz} leads to $L_{\mathrm{rem}} = 72$~cm$^{-2}$ and 36~cm$^{-2}$ for the D7 and D10 channels, respectively. What the experiment can access, however, is the \emph{variation} of $L$ in the quiescent state between one decay and another, which constitutes a measurement of the remanent vorticity above the unmeasurable $L_{\mathrm{rem}}$ floor. We calculate this quantity as
\begin{equation}
\label{eq:LR}
L_{R} = L_{\mathrm{start}} - L_{\mathrm{end}},
\end{equation}
where this difference is obtained from the change in average amplitude of the second-sound signal averaged over a period of 20~s in the quiescent states before and after the imposition of a flow. We have studied the distribution of $L_{R}$ in the batch of 150 decays across all parameter space. An example for D7, $T = 1.65$~K and L5 is given in the inset of Fig.~\ref{fig:statistics}, showing roughly a Gaussian distribution centered around zero; this distribution is common to all batches. Notice that $L_R$ can be positive or negative, and that the extent of its variation cannot be accounted for by changes in helium temperature from one decay to another (which are too small for this effect); therefore we attribute this effect to varying remanent vortex line density in the sample. The standard deviation of this distribution, $\mathrm{SD}(L_{R})$, is given in the main plot of Fig.~\ref{fig:statistics} as a function of $T$ and $L_0$ for the D7 channel, D10 being similar. From this we learn that $\mathrm{SD}(L_{R})$ is not correlated to $L_0$ and that with the exception of the points at $T = 2.1$~K where measurements are more difficult, it varies from about 100~cm$^{-2}$ to 300~cm$^{-2}$. We guess therefore that an upper limit on the absolute remanent vortex line density in the sample is roughly $\mathrm{SD}(L_{R})$.

We do not know precisely what effect remanent vortices might have on the rate of decay of a vortex tangle, compared to a decay where vortex lines are totally annihilated, a condition which, as we have explained, cannot be achieved experimentally due to pinning. However we have indication that when the counterflow heat current is reduced not to zero, but to a finite subcritical value, the decay shape is somewhat affected (Fig.\ref{fig:comp1}) -- this suggests that the decay rate may be affected by the amount of residual vortices. A detailed study of these effects was carried out for thermal counterflow~\cite{vinen_transient} and we plan a similar approach for mechanical superflow.

In addition to these effects a further complication for interpretation arises at low densities: for $T < 2$~K our averaged estimated residual density $\mathrm{SD}(L_{R}) \approx 250$~cm$^{-2}$ corresponds to a ratio of line separation to channel width of about 0.1 in D7, at which wall effects might well start to be important. For these combined reasons, we think it wise to consider our results only up to times for which $L \gtrapprox \mathrm{SD}(L_{R})$, despite the fact that our ensemble-average curves resolve a longer decay process. For example, for the decay in Fig.~\ref{fig:ensemble} for which $\mathrm{SD}(L_{R}) \approx 100$~m$^{-2}$  we would limit our consideration up to about $t = 90$~s. This is taken into account in the rest of the article.

\section{Experimental results}
\label{sec:results}

\begin{figure*}
\includegraphics[width = 0.7\linewidth]{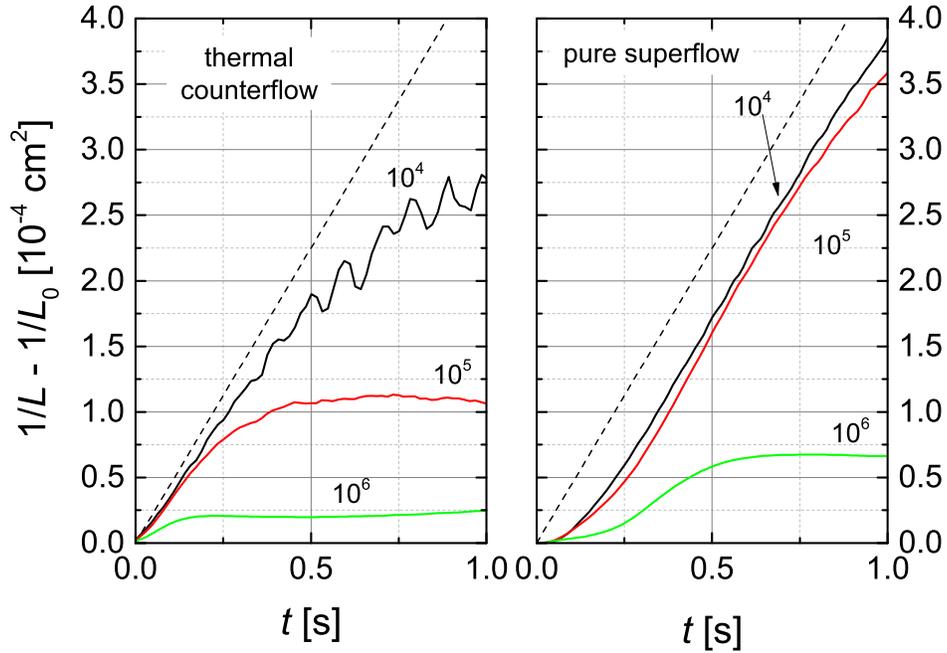}%
\caption{(Color online) Comparison of decays observed for thermal counterflow with those observed for bellows-driven superflow for different initial line densities at very small times for $T = 1.65$~K. Plots of $1/L-1/L_0$ against $t$ according to Eq.~\eqref{eq:Vinen_decay_solution_inv}. The Vinen decay of a random tangle is observed for a very short transient and is followed more closely by mechanical superflow than thermal counterflow, and, in both cases, by decays originating from lower densities. (The $T = 1.45$~K data are in SM)}
\label{fig:comp3}
\end{figure*}

\begin{figure}
\includegraphics[width = 1\linewidth]{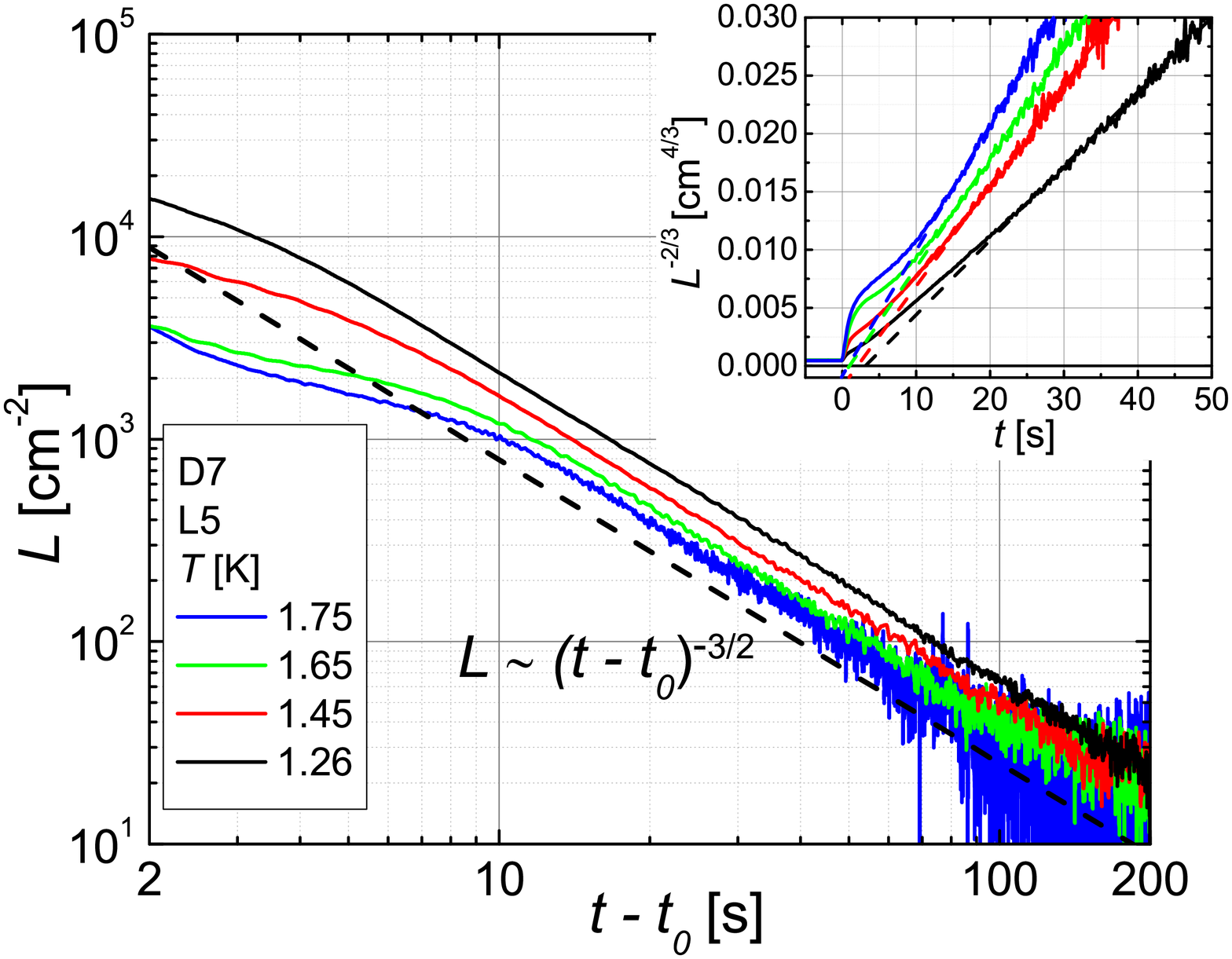}%
\caption{(Color online) Superflow decay in the D7 channel from initial line density $L_0 = 10^5$~cm$^{-2}$ at different temperatures. In the main plot $L$ and $t$ are plotted in logarithmic axis, demonstrating that superflow decays with the power law $L \propto (t - t_0)^{-3/2}$ (dashed line) for a very large fraction of time, except for the first few seconds, according to the quasi-classical single fluid decay model in Eq.~\eqref{eq:classical_limit}. The x-axis requires the subtraction of the virtual time origin $t_0$ obtained from the inset, as described in the text. Similar plots from different initial density and for the D10 channel are in SM and confirm the same conclusions.}
\label{fig:sf_long_t_L5_D7}
\end{figure}

\begin{figure*}
\includegraphics[width = 0.7\linewidth]{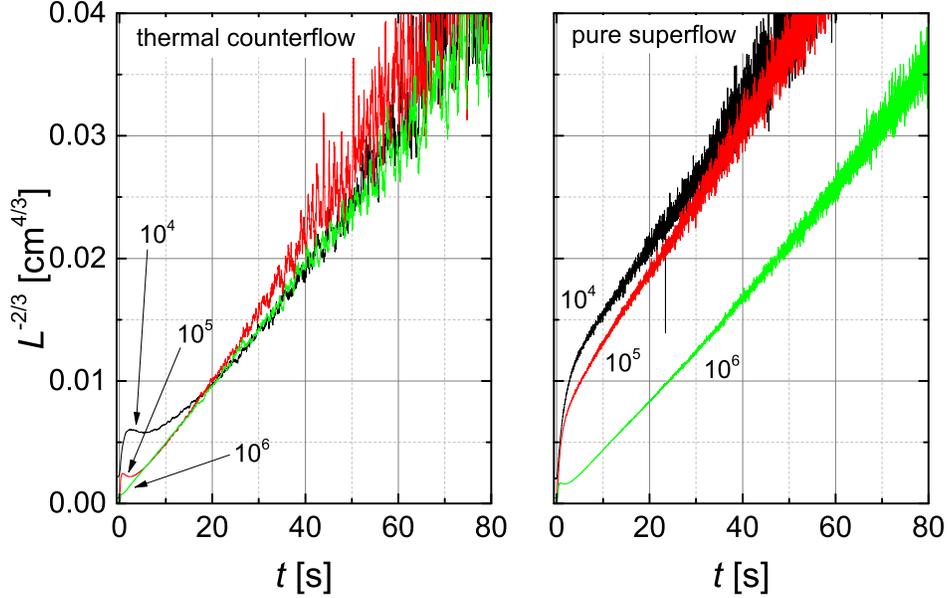}%
\caption{(Color online). Comparison of decays observed for thermal counterflow with those observed for bellows-driven superflow for different initial line densities at $T = 1.65$~K in D10. Plots of $L^{-2/3}$ against $t$ as according to Eq.\eqref{eq:classical_limit_inverse}. Linear behavior corresponds to the quasi-classical $L \propto t^{-3/2}$, observed in both flows. The $T = 1.45$~K version is in SM.}
\label{fig:comp2}
\end{figure*}

An overview of the first 10~s of all our experimentally observed superflow decays in the D7 and D10 channels is shown in Figs.~\ref{fig:sf_D7_short} and \ref{fig:sf_D10_short} respectively. Each panel groups decays from the same initial density and different temperature. These plots serve to demonstrate the main features of the decays at a glance; details on shorter and longer time-ranges will be presented in due course. An overview of the full time range in log-log coordinates is available for D7 in the Supplemental Material (SM), with D10 being similar. Each decay shown here in Figs.~\ref{fig:sf_D7_short} and \ref{fig:sf_D10_short}  displays an initial fast rate (see also Fig.~\ref{fig:comp3} for details) which expires within roughly 0.5 and 2~s depending on conditions, followed by a slower process which continues for a longer time (see also Figs.~\ref{fig:sf_long_t_L5_D7} and \ref{fig:comp2} for details).  For short time behavior, we notice that qualitatively but systematically the ratio of $L_0$ to $L_{x}$ at which the fast process changes to a slow process decreases with $L_0$ for fixed $T$, and increases with $T$ for fixed $L_0$, in both channels. The time at which $L_x$ occurs  increases systematically with decreasing $L_0$ for fixed $T$ and increases with $T$ for fixed $L_0$.

\begin{figure*}
\includegraphics[width = 0.7\linewidth]{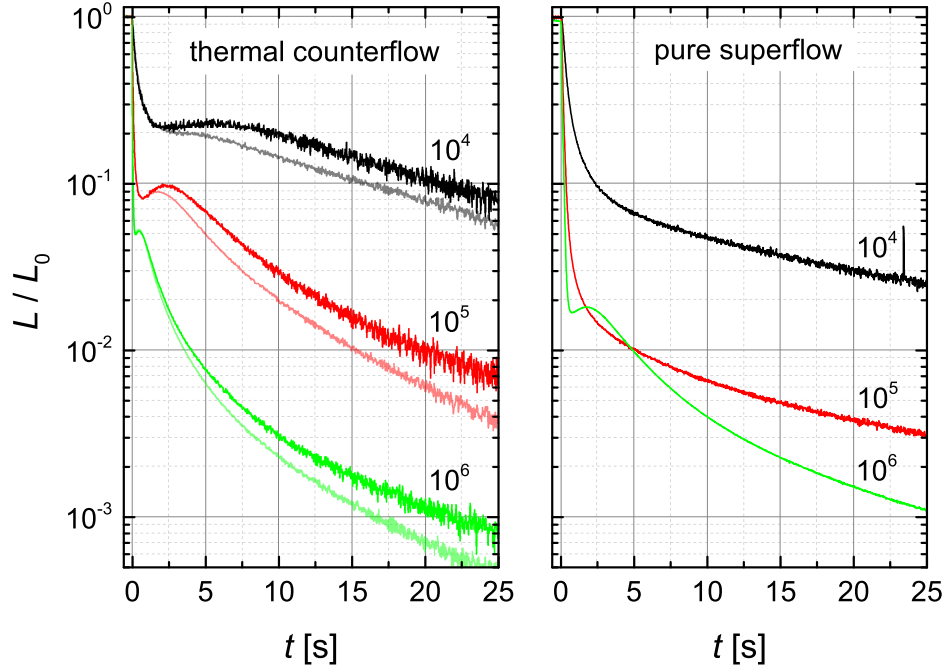}%
\caption{(Color online) Comparison of decays observed for thermal counterflow with those observed for bellows-driven superflow for different initial line densities, at the same temperature, $T= 1.65$~K, and in the same channel, D10. Line densities are normalized to the initial line densities. For counterflow, lighter lines correspond to decays when the heat current was reduced to a sub-critical value of 10~mW. The temporary increase in vortex line density during decay (the ``bump'') is observed systematically in thermal counterflow, but only in the D10 channel and at high density in mechanical superflow. Similar plots at $T = 1.45$~K are in the Supplemental Material (SM), including also the thermal counterflow decay at $T = 2.10$~K which does not have a counterpart in superflow for the same channel size.}
\label{fig:comp1}
\end{figure*}

In D10-L6 only (Fig.~\ref{fig:sf_D10_short} - left panel), the fast and slow regimes are joined by an intermediate one, with an inversion of the decay rate, i.e. a change of sing of $dL/dt$. In D10-L5 at $T = 1.25$~K, instead, we observe only a slow down of the decay rate without an increase in $L$, but with the presence of a point of inflexion across which $d^2L/dt^2$ changes sign. We shall refer henceforth to these features as a ``bump'' in the decay curve. For simplicity we shall call ``bump'' both the increase in $L$ during the decay and the presence of a point of inflexion, and we shall be more specific when needed. The bump, observed only in these few circumstances in superflow, is instead seen always in counterflow in the stricter sense of non-monotonic behaviour, as demonstrated in Fig.~\ref{fig:comp1} and related Supplemental Material (SM). The existence of a non-monotonic bump in counterflow was already detected in earlier experiments~\cite{prague_counter1, prague_counter2, barenghi_skrbek_rev} and here, within the range of investigated parameters, we confirm it. There were however instances in past experiments, including the once just cited, when a point of inflexion was observed instead of non-monotonic behaviour, as in the first experiments on the decaying thermal counterflow by Vinen~\cite{vinen_transient} or more recently in the Prague decays~\cite{prague_counter1, prague_counter2}, for decays originating from steady-state counterflow generated by larger heat fluxes than reached here. What Fig.~\ref{fig:comp1} adds to previous studies is high statistics for counterflow (ensemble average of 150 decays instead of single curve, i.e. the same standards as for our superflow) and, especially the fact that we can compare superflow and counterflow strictly under the same conditions: channel, temperature, and initial vortex line density.

Fig.~\ref{fig:comp1} shows that the bump in counterflow is broadened with decreasing $L_0$, and that in superflow it occurs only for L6,  although the details of the decay remain different in the two cases. The lighter curves in the left panel are decays when the heat current is reduced, not to zero, but to a small value (10 mW) below the critical value for transition to quantum turbulence. The idea here was to observe if leaving a subcritical heat current in the channel may help in reducing the level of residual vortex lines, ``washing them away'', in the spirit of the systematic study in Ref.~\cite{vinen_transient}. Whilst we did not observe a change in the residual second-sound attenuation at the end of the decay process, we did measure differences during the rest of the decay, as shown. At late times the line density is lower in case of non-zero heat current, and the ``bump'' is somewhat reduced.

Let us now focus on the fast initial decay regime observed at short times, emphasized in Fig.~\ref{fig:comp3} for counterflow and superflow in D10 at $T = 1.65$~K. In the figure we compare this decay regime with the prediction of the Vinen model in Eq.~\eqref{eq:Vinen_decay_solution} relating to the decay of a fully random tangle unbounded by walls, by recasting the equation as follows:
\begin{equation}
\frac{1}{L} - \frac{1}{L_0}=  \Big(\frac{\chi_2  \kappa}{2 \pi}\Big) t,
\label{eq:Vinen_decay_solution_inv}
\end{equation}
where the quantity $\kappa \chi_2 / 2 \pi$ is given in Table~\ref{tab:beta}. We notice that the Vinen decay is not generally observed, and when it exists, is followed for only a very short time, at most 1~s. The pure superflow is more Vinen-like than the counterflow, and departures increase with increasing $L_0$ in both cases. The situation is similar at other temperatures, with departure from Vinen behavior increasing with decreasing temperature and increasing initial density (data at $T = 1.45$~K are provided as SM). In Fig.~\ref{fig:comp3} we notice also that the decay does not start abruptly as predicted by Eq.~\eqref{eq:Vinen_decay_solution}, but there is some rounding immediately after $t = 0$. This rounding is more pronounced in superflow than in counterflow, lasting at most some 200 ms, with the tendency to increase with initial line density. This latter fact leads us to favour the explanation that the rounding may be an instrumental effect. The switching off of counterflow is controlled fully electronically, whilst the bellows is a mechanical system which may introduce secondary lagging effects. Although the bellows motor encoder at room temperature does show that the system comes to rest to within 10~ms (Fig.~\ref{fig:rawdata}), the actual flow may not stop abruptly, for reasons such as finite compressibility, the finite time for expiration of pressure gradients (which would increase with bellows velocity as observed), the expiration of thermal gradients occurring because of the presence of superleaks causing an increase of temperature in the bellows and a decrease in the channel due to change in superfluid/normal density ratio (see Ref.~\cite{babuin_steady_superflow} for more detailed discussion), and so on. In principle however one should not exclude the possibility that the flow actually stops at $t = 0$ and therefore the rounding would be explained by some incompleteness of the model in Eq.~\eqref{eq:Vinen_decay_solution}. At any rate, our time resolution of $\approx 16$~ms, limited by the intrinsic physics of second-sound resonance, does not allow  us to study this physical process in greater depth. We therefore concentrate on the decay process after the first, say, 100 ms have elapsed.

\begin{table}
\centering
\begin{tabular}{|c|c|}
\hline
$T$[K] & $\kappa \chi_2 / 2 \pi$ [cm$^2$s$^{-1}$] \\
\hline
1.25 & $2.2 \times 10^{-4}$\\
1.45 & $3.1 \times 10^{-4}$\\
1.65 & $4.5 \times 10^{-4}$\\
1.75 & $5.3 \times 10^{-4}$\\
2.10 & $9.3 \times 10^{-4}$\\
\hline
\end{tabular}
\caption{Values of $\kappa \chi_2 / 2 \pi$ obtained averaging the values computed by Schwarz~\cite{schwarz1988} and those given by the theory of Vinen and Niemela~\cite{vinen_niemela_rev}. This quantity occurs in Eq.~\eqref{eq:Vinen_decay_solution_inv} which is plotted in Fig.~\ref{fig:comp3}.}
\label{tab:beta}
\end{table}

The situation at late times is summarized in Fig.~\ref{fig:sf_long_t_L5_D7}, for the case of superflow in D7, with starting density L5 at different temperatures (D10 version in SM). Although we know from section~\ref{subsec:remanent} that we should handle with care the decay process when $L$ becomes comparable with or smaller than its mean remanent value, in this figure we demonstrate that a single power law of the form $L \propto (t-t_o)^{-3/2}$ represents the data from the first few seconds to the end of the range. This is the behavior predicted by the quasi-classical model in Eq.~\eqref{eq:classical_limit}.

\begin{figure}
\includegraphics[width = 1.0\linewidth]{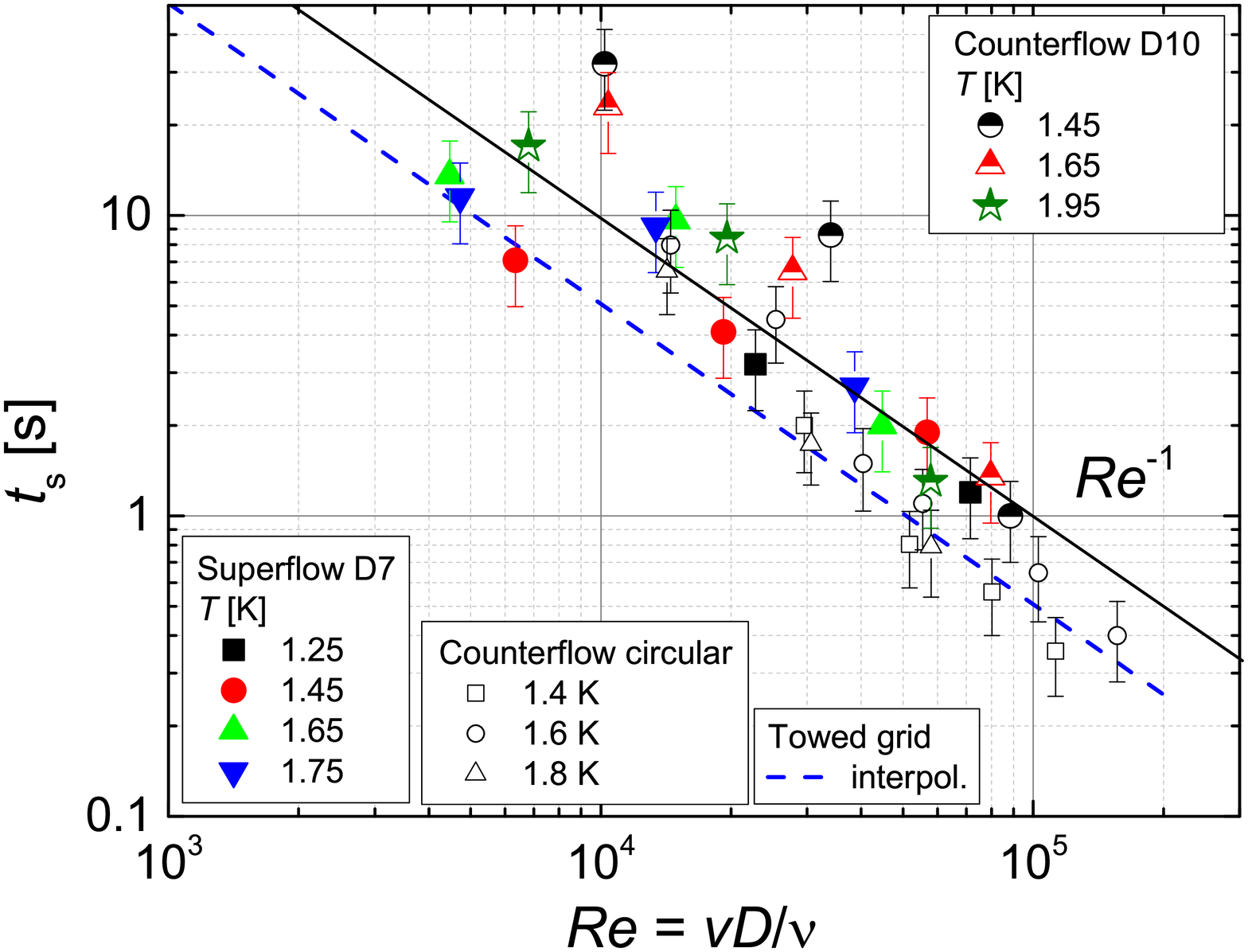}%
\caption{(Color online) The time for the onset of $L \propto t^{-3/2}$ behavior (saturation time, see text) as a function of the flow's Reynolds number. The present experiments are compared with an older Prague thermal counterflow experiment in a round channel~\cite{prague_counter1} and with the Oregon towed grid experiments (the dashed-line is an interpolation of their experimental points) which measures the turbulence decay behind a moving grid~\cite{skrbek_oregon}.}
\label{fig:saturation}
\end{figure}

Since Eq.~\eqref{eq:classical_limit} has a virtual time-origin $t_0$, this must be subtracted from the time axis in a logarithmic plot. To estimate it we recast Eq.~\eqref{eq:classical_limit} as follows
\begin{equation}
\frac{1}{L^{2/3}} = \frac{(2 \pi)^{2/3} \kappa^{2/3} \nu'^{1/3}}{3 C D ^{2/3}}(t - t_0),
\label{eq:classical_limit_inverse}
\end{equation}
and obtain $t_0$ as the intercept of the linear part of the plot with the time axis, as demonstrated in the inset of Fig.~\ref{fig:sf_long_t_L5_D7} by dashed lines. The time at which the $t^{-3/2}$ behavior onsets, i.e. the ``saturation time'' $t_s$ (``saturation'' refers to the attainment of the condition of large eddies reaching their maximum size limited by the channel width $D$, required in deriving Eq.~\eqref{eq:classical_limit}), is plotted for different flows in Fig.~\ref{fig:saturation} against the Reynolds number, defined with the mean velocity $v$, the channel width $D$, and the kinematic viscosity of the normal component $\nu$, following Ref.~\cite{prague_counter1}. Allowing for some scatter, all experiments produce roughly a scaling $t_s \propto Re^{-1}$.

We continue the analysis of the late time behavior by contrasting superflow and counterflow in Fig.~\ref{fig:comp2}, for the same temperature, $T = 1.65$~K and channel, D10, observing that in both cases there is linear range as predicted by Eq.~\eqref{eq:classical_limit_inverse} extending as far as 80 s-- the similar $T = 1.45$~K case is in SM. This equation also predicts that (i) curves from different initial densities should collapse: our data confirms that decays from different $L_0$ have very similar slope in the $t^{-3/2}$ regime, but the collapse is observed only in counterflow and not in superflow. (ii) The slope should scale as $D^{-2/3}$: despite the fact that we have only two channel sizes, we checked this prediction and found that experimental slopes scale by a substantially larger extent than expected. In these two respects therefore our results differ from the prediction.

\begin{figure}
\includegraphics[width = 1.0\linewidth]{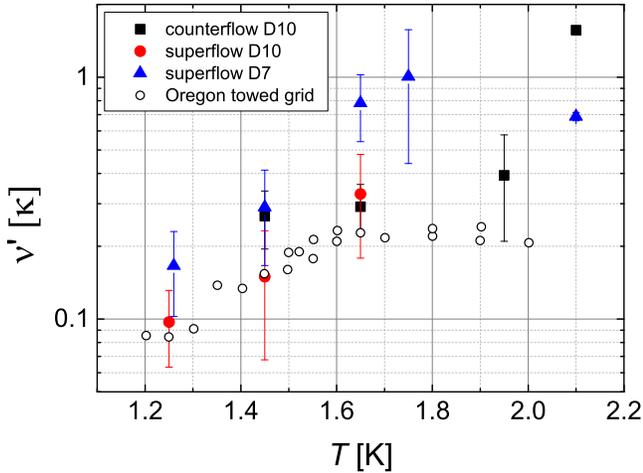}%
\caption{(Color online) The effective viscosity of the ``quasi-classical'' model (Eq.~\eqref{eq:classical_limit}) as a function of temperature. For our three experiments, at each given temperature, $\nu'$ has been averaged from decays of different initial density (L4, L5 and L6) , resulting in the shown error bars. For comparison we add also data from the Oregon towed grid experiment~\cite{niemela_grid}.}
\label{fig:viscosity}
\end{figure}

Fitting Eq.~\eqref{eq:classical_limit_inverse} to the linear part of the curves in Fig.~\ref{fig:comp2} allows us to extract the effective viscosity $\nu'$, as is customarily done~\cite{skrbek_sreeni_10ch, Walmsley_rev, prague_eff_kin}. In Fig.~\ref{fig:viscosity}, together with the data from the decay of turbulence past the towed grid~\cite{niemela_grid}, we thus plot $\nu'(T)$ for all our flows. The effective viscosity measured from decays starting from different vortex line density $L_0$ is found to slightly vary, but in a manner uncorrelated with $L_0$. In any case, we do not expect $\nu'$ to be dependent on $L_0$ because we understand $\nu'$ to be a robust property of the flow independent of flow details, as it is shown by consistent values coming from different decay experiments~\cite{prague_eff_kin} and also from steady-flows, hence from an entirely independent approach~\cite{babuin_viscosity}.
For these reasons, we think it justified to average $\nu'$ originating from different $L_0$ -- which is reflected in the error bars in Fig.~\ref{fig:viscosity}. We notice that, despite the data from the three experiments at the same temperature being roughly consistent within error bars, the superflow data for D10 lie systematically lower than those for D7, which puts pressure on the understanding that $\nu'$ should not depend on flow details. At any rate, as we have discussed in our works which include determination of effective viscosity~\cite{babuin_viscosity, babuinJLTP}, the absolute value of $\nu'$ obtained from decay measurements can have uncertainties up to a factor of 4, depending on how strictly the assumption on saturation of the large eddies size attains. Accurate experimental determination of $\nu'$ to better than a factor of 4 remains therefore still a challenge.

\section{Discussion}
\label{sec:discussion}

All our experimentally observed decays, both in superflow and counterflow, have an initial regime when the line density drops rapidly (for sufficiently low density it approaches for a short time the Vinen law $L \propto t^{-1}$, especially in superflow), and a final slower regime of the form $L \propto t^{-3/2}$. The first, fast regime, lasts at most 1 s (Fig.~\ref{fig:comp3}) and is responsible for the loss of a large fraction of vortex lines, even up to about 99\% (see, e.g., the $T = 1.75$~K curve in Fig.~\ref{fig:sf_D7_short}, left panel). The fact that the decay rate increases with increasing temperature for fixed initial density may be due to rising value of the prefactor $\kappa \chi_2 / 2 \pi$ in Eq.~\eqref{eq:Vinen_decay} with temperature, as shown in Table~\ref{tab:beta}, which in turn  must be related to the increase in mutual friction. The second, slow regime, lasts for the majority of the decay process and causes the loss of the remaining vortex lines (except the remanent ones).

 We have compared our observations with two available analytic models: (i) the Vinen equation for a random tangle - and found it to apply in a limited regime and (ii) the quasi-classical model for the decay of large eddies - and found it to apply more generally although not fulfilling all predictions.

 Our results generally confirm the understanding~\cite{Walmsley_rev, baggaley_PRL, roche_model} that a general quantum turbulence tangle consists of an random system of vortex lines which decays fast by mutual annihilation of lines, a fraction of which is organized in bundles giving rise rise to eddies of various sizes, up to sizes comparable to the channel width, which decay more slowly. Alternatively,  we can imagine there being a more spatially homogeneous regime with an energy spectrum that is Kolmogorov in form except for an initially enhanced energy at wave numbers close to the inverse vortex line spacing;  i.e a situation in which the density of vortex lines is initially larger than is necessary for the dissipation of energy,  given by Eq.~(\ref{eq:diss}), at the rate required to match the flow of energy down the Kolmogorov cascade.   Our experiment cannot establish unequivocally whether these eddies exist already in the steady-state, although evidence that they do exist comes from visualization of counterflow turbulence both from tracing solid particles~\cite{MarcoPRB} and imaging the normal component~\cite{guo2010, guo2015}.

What is not predicted by the existing analytic models is the occurrence of the ``bump'', i.e., the change of sign in $dL/dt$ or $d^2L/dt^2$ between the fast and slow regimes, which we have observed always in thermal counterflow and as an exception in superflow. The second-sound attenuation can become increased during decay for one, or for a combination of two reasons: (i) the line length stays approximately constant but it is spatially rearranged so that the second-sound ``sees'' a greater fraction (recall that second-sound is attenuated only by the projection of vortex lines onto a plane perpendicular to second-sound's propagation direction), or (ii) the spatial orientation stays fixed but the line length increases.

The possibility of (i) has been confirmed numerically by Barenghi \emph{et al.}~\cite{barenghi_skrbek_depol, barenghi_skrbek_rev} for spatially rearranged random tangles, but although the simulation gave a qualitative result in agreement with observations, the vortex line density in the simulation was about an order of magnitude below the lowest available experimentally. We note nevertheless that the observed height of the bump is never greater than can be accounted for by this mechanism (33\%).  On the other hand, this effect was not observed in the simulations of Mineda and coworkers~\cite{mineda},  based on more realistic line densities. However,  neither of these simulations takes account of the possibility that the vortex tangles with which we are dealing are polarised in such a way that large scale eddy motion is superimposed on the random tangle.

Option (ii) is also physically possible, and to build an argument for it we note that in our measurements the bump occurs always in thermal counterflow and only at high steady-state velocities in superflow. This suggests that the dynamical state of the normal component (laminar, unstable or turbulent) might be relevant to the existence or not of the bump, since the average velocity of the normal component relative to the walls varies significantly in the two systems: it can be rather large in the case of thermal counterflow, but it is nominally zero in the case of superflow. We note however that, as suggested in Ref.~\cite{babuin_steady_superflow}, from considerations on the scaling of critical velocity with channel width from different experiments, the normal component is probably not at rest in superflow in a large channel, but is set in motion by mutual friction, the spatially-averaged velocity remaining zero. Nevertheless this motion may be relatively slow in comparison with that in thermal counterflow, except at large superflow velocities. But just how would the motion of the normal component cause the decay inversion? The recent results from the Tallahassee group~\cite{guo2015} where the normal component is tracked by the excimer molecules, give evidence that the normal component at a sufficiently high heat current becomes turbulent (specifically at $T = 1.83$~K for a heat current above $q_c = 80$~mW/cm$^2$ at which $L_0 \approx 10^4$~cm$^{-2}$). The second order structure functions calculated from the normal fluid turbulent velocity fluctuations extend over a fairly large range of lengths scales, up to a sizeable fraction of the channel size. The structure function can be related to a turbulent energy spectrum of the form $E(k) \propto k^{-2}$ in the steady-state (where $k$ is the wavenumber).  During the decay the spectrum changes gradually, within about 3 seconds, into the Kolmogorov form $E(k) \propto k^{-5/3}$. It seems possible that the bump is associated with this evolution. We are currently exploring this idea with a model for the temporal evolution of the energy spectrum, and this work will be reported in a future publication. Further experimental evidence that the bump is associated to transition to turbulence in the normal component would come by testing thermal counterflow at heat currents below the transition. In our system this would mean to study decays from initial density sufficiently below $L_0 = 10^4$~cm$^{-2}$, which constitutes a direction for future work.

\section{Conclusions}
\label{sec:conclusions}

We have presented a comprehensive picture of the temporal decay of vortex line density in quantum turbulence produced by mechanically driven superflow through two square channels of 7 and 10~mm side. We have covered a broad parameter space in temperature ($1.25 \leq T \leq 2.10$~K) and in steady-state vortex line density ($10^4  \leq L \leq  10^6$~cm$^{-2}$). Additionally, at $T = 1.45$~K and $T = 1.65$~K and for all the same initial densities, we have provided, for the first time, a direct comparison of mechanical superflow and thermal counterflow, the latter performed in the same 10~mm wide channel used for superflow and under exactly the same experimental conditions. This, together with enhanced accuracy achieved thanks to ensemble averaging of up to 150 individual decays placed us in a strong position to compare these flows.

In an unbound system superflow and counterflow ought to display identical physics since they are related by Galilean invariance. In practice, the presence of channel walls will change the physics, at least for the normal component which must acquire a profile due to viscous drag with the walls. This has been indeed confirmed experimentally by visualizing the normal fluid flow profile using helium excimer molecules, observed to turn from laminar to turbulent as heat current increases~\cite{guo2015}. Additionally, we know from numerical simulations~\cite{BaggaleyLaizet, yui, LvovPRBR}  that the normal fluid profile induces inhomogeneity in the distribution of vortex lines across the channel width, with the density being enhanced in case of turbulent normal fluid profile. Our experiments with steady-state superflow have shown that when comparing with other counterflow experiments with channels of different size~\cite{babuin_steady_superflow} and also when measuring in our own superflow channel adapted for counterflow~\cite{Babuin_PF}, the line density for a given relative velocity is essentially insensitive to the normal component net flow. This is not so for the decay.

The decay of these two flows is similar in that both display an initial fast process where most of the line density is lost which for sufficiently low density has the form $ L\propto t^{-1}$, and a subsequent slow $ L\propto t^{-3/2}$ process where the rest of the tangle decays. These two processes have been associated respectively with the decay of the randomized and polarized components of the tangle. The key difference however is that in counterflow we invariably observed an inversion of the decay rate between the two regimes, which we observed only at high steady-state velocities in the wider channel in superflow. This fact, firmly established by high-statistics measurements indicates that the dynamical state of the normal component in the steady-state (no net flow through the channel in superflow and turbulent pipe flow in counterflow) has consequences for the decay of the tangle. We have speculated about the reasons, but further work is required if these reasons are to command confidence. We hope that this comprehensive set of experimental data describing the vortex tangle decay in superflow and counterflow will stimulate the development of a still missing full theory of counterflow turbulence.

\begin{acknowledgments}
We thank M. Stammeier for his help in the early stages of this experimental research. This work was supported by GA\v{C}R GA14-02005S.
\end{acknowledgments}

\bibliographystyle{apsrev4-1}
\bibliography{qtbib}

\end{document}